\documentclass[twocolumn,english,prl]{revtex4-1}
\usepackage[T1]{fontenc}
\usepackage[utf8]{inputenc}
\usepackage{color}
\usepackage{prettyref}
\usepackage{amsmath}
\usepackage{amssymb}
\usepackage{graphicx}

\makeatletter

\@ifundefined{textcolor}{}
{%
 \definecolor{BLACK}{gray}{0}
 \definecolor{WHITE}{gray}{1}
 \definecolor{RED}{rgb}{1,0,0}
 \definecolor{GREEN}{rgb}{0,1,0}
 \definecolor{BLUE}{rgb}{0,0,1}
 \definecolor{CYAN}{cmyk}{1,0,0,0}
 \definecolor{MAGENTA}{cmyk}{0,1,0,0}
 \definecolor{YELLOW}{cmyk}{0,0,1,0}
}

\makeatother

\usepackage{babel}
\begin{document}

\title{Real time measurement of the emergence of superconducting order in
 a high temperature superconductor.}

\author{I. Madan$^{1}$, P. Kusar$^{1}$, V. V. Baranov$^{1}$, M. Lu-Dac$^{1}$,
V. V. Kabanov$^{1}$, T. Mertelj$^{1}$, and D. Mihailovic$^{1,2}$}

\affiliation{$^{1}$Complex Matter Department, Jozef Stefan Institute,}

\affiliation{$^{2}$CENN Nanocentre, Jamova 39, 1000 Ljubljana, Slovenia}

\date{\today}
\begin{abstract}
Systems which rapidly evolve through symmetry-breaking transitions
on timescales comparable to the fluctuation timescale of the single-particle
excitations may behave very differently than under controlled near-ergodic
conditions. A real-time investigation with high temporal resolution
may reveal new insights into the ordering through the transition that
are not available in static experiments. We present an investigation
of the system trajectory through a normal-to-superconductor transition
in a prototype high-temperature superconducting cuprate in which such
a situation occurs. Using a multiple pulse femtosecond spectroscopy
technique we measure the system trajectory and time-evolution of the
single-particle excitations through the transition in La$_{1.9}$Sr$_{0.1}$CuO$_{4}$
and compare the data to a simulation based on time-dependent Ginzburg-Landau
theory, using laser excitation fluence as an adjustable parameter
controlling the quench conditions in both experiment and theory. The
comparison reveals the presence of significant superconducting fluctuations
which precede the transition on short timescales. By including superconducting
fluctuations as a seed for the growth of superconducting order we
can obtain a satisfactory agreement of the theory with the experiment.
Remarkably, the pseudogap excitations apparently play no role in this
process.
\end{abstract}
\maketitle
The study of the time evolution of complex systems through symmetry
breaking transitions (SBT) is of great fundamental interest in different
areas of physics\cite{Bunkov2000,Higgs1966,Volovik2003}. An SBT of
particular general interest is the normal-to-superconducting $(N\rightarrow S)$
state transition in which a Lorentz non-invariant system breaks gauge
invariance.\cite{Varma2002} By studying the $N\rightarrow S$ transition
in time-evolving systems, rather than by slowly varying the temperature
through the transition, one can in principle gain new information
on the dynamical behavior of elementary excitations which lead to
the formation of a superconducting condensate and the collective ordering
behavior, leading to new insights into non-ergodic phenomena of collectively
ordered systems as well as the mechanism of superconductivity.\emph{
}Particularly, ergodicity breaking  in rapidly evolving systems leads
to the appearance of topological defects (vortices). 

The description of the dynamical behavior of the gauge non-invariant
systems is given in the time-dependent Ginzburg-Landau theory (TDGL
theory). It has been first applied to the problem of non-equilibrium
phase transitions by Kibble and Zurek who considered the appearance
of topological defects throughout the transition.\cite{Kibble1997,Zurek1996}
The Kibble-Zurek description has been indirectly confirmed to be correct
by static experiments in which trapped vortices  were studied.\cite{Monaco2006,Golubchik2010}
In this paper, beyond previous static studies, we study \emph{real-time
evolution} of the superconducting order in the non-equilibrium phase
transition. We investigate the applicability of TDGL theory to the
phase transition problem and provide a minimal formulation sufficient
to describe the data.

The paper is organized in the following way: we first introduce the
problem of a non-homogeneous non-equilibrium phase transition, then
present the three-pulse technique we employed and describe acquired
data. In the second half of the paper we present the numerical simulations
of the $S\rightarrow N$ and $N\rightarrow S$ transitions with TDGL
theory, using different formulation of the problem.

\subsection*{Laser induced nonequilibrium phase transition}

An experimental realization of nonequilibrium conditions requires
the inverse of the cooling rate - the quench time $\tau_{\mathrm{q}}$,
to be comparable to the intrinsic collective system relaxation time
$\tau_{\mathrm{GL}}=\pi$$\hbar/8k(T-T_{\mathrm{c}})$$\simeq10^{-13}-10^{-12}$
s.\cite{Volovik2003,Zurek1985,Schmid1975,Schmid1966} So far the quench
was physically limited to the ns timescale by heat diffusion processes
or the duration of the optical pulse used for driving the transition.\cite{Golubchik2010}

With femtosecond optical spectroscopy, nonequilibrium regime of the
phase transition as well as measurements of the critical region in
real time on the timescale of $\tau_{\mathrm{GL}}$ become accessible.
By properly adjusting pulse energy the limitations on the quench time
set by heat diffusion processes can be overcome: for moderate fluences
the electronic subsystem gets highly perturbed \cite{Kusar2008,Giannetti2009a,Coslovich2011}
while the lattice remains only weakly excited. In this case the cooling
rate is defined by the energy exchange between electronic and lattice
subsystems, which typically occurs on the sub-ps timescale\cite{Gadermaier2010},
which is much faster than heat diffusion. 

Optical experiments are intrinsically inhomogeneous due to finite
light penetration depth $\lambda_{p}$. This affects not only the
data analysis but also the physics of the transition. Due to the exponential
depth-distribution of absorbed energy the superconducting condensate
is destroyed only up to a certain depth. This results in a sharp boundary
between the \emph{N} and \emph{S} states. After the quench the boundary
propagates towards the surface and is expected to reach it on a timescale
$\tau_{\psi}\sim\lambda_{\mathrm{p}}/v_{\psi}\sim\lambda_{\mathrm{p}}\tau_{\mathrm{GL}}/\xi_{\bot}\sim10^{3}\,\tau_{\mathrm{GL}}$,
where $v_{\psi}$ is the velocity of the $S/N$ boundary%
\footnote{The 1-D time dependent GL equation $\tau_{\mathrm{GL}}\dot{\psi}=\xi^{2}\psi''+(1-T/T_{\mathrm{\mathrm{c}}})\psi-\psi^{3}$
has a propagating soliton solution $\psi\propto\tanh[(v_{\psi}t-z)/w]-1$
where $w=2\sqrt{2}\xi/\sqrt{1-T/T_{\mathrm{c}}}$ and $v_{\psi}=3\xi\sqrt{1-T/T_{\mathrm{c}}}/\sqrt{2}\tau_{\mathrm{GL}}.$
Taking $\xi\sim\xi_{\bot}\sim0.2$ nm and $\tau_{\mathrm{GL}}\sim100$
fs gives $v_{\psi}<5$ nm/ps.\label{fn:SOliton}%
} and $\xi_{\bot}$ is the out-of-plane\emph{ S} coherence length.
Though the boundary propagation is relatively slow, compared to $\tau_{\mathrm{GL}}$,
the physics of the transition depends on how it relates to the propagation
of the temperature front, which is defined by the quench conditions.
Two regimes are possible: the temperature front propagation velocity
$v_{\mathrm{T}}$ can be either larger (rapid quench) or smaller (slow
quench) than the characteristic critical value $v_{\mathrm{crit}}\approx v_{\psi}\sqrt[4]{\tau_{\mathrm{GL}}/\tau_{\mathrm{q}}}\sim10^{5}$
cm/s .\cite{Kibble1997} In the rapid quench limit when $v_{\mathrm{T}}>v_{\mathrm{crit}}$,
the normal region between the temperature front and $S/N$ boundary
is supercooled and the order parameter grows from fluctuations. In
this case one can expect vortex formation according to the Kibble-Zurek
(KZ) mechanism. In the slow quench limit $(v_{\mathrm{T}}<v_{\mathrm{crit}}$)
the condensate forms instantaneously in the wake of the temperature
front so that the phase of the order parameter is defined by the bulk
value and vortex formation becomes suppressed\cite{Aranson1999,Kibble1997}.

As we shall see, both cases are accessible in our experiments via
variable laser fluence: At low fluences, only the electrons are heated
above $T_{\mathrm{c}}$. They cool rapidly through $T_{\mathrm{c}}$,
so the quench rate $\gamma_{\mathrm{\mathrm{q}}}=\left(dT/dt\right)/T_{\mathrm{c}}$
is fast\cite{Gadermaier2010}. With large fluences, the lattice is
heated above $T_{\mathrm{c}}$. It's cooling is defined by the heat
diffusion so the quench rate is much slower.

\begin{figure}[t]
\begin{centering}
\begin{minipage}[t]{1\columnwidth}%
\begin{center}
\includegraphics[width=1\columnwidth]{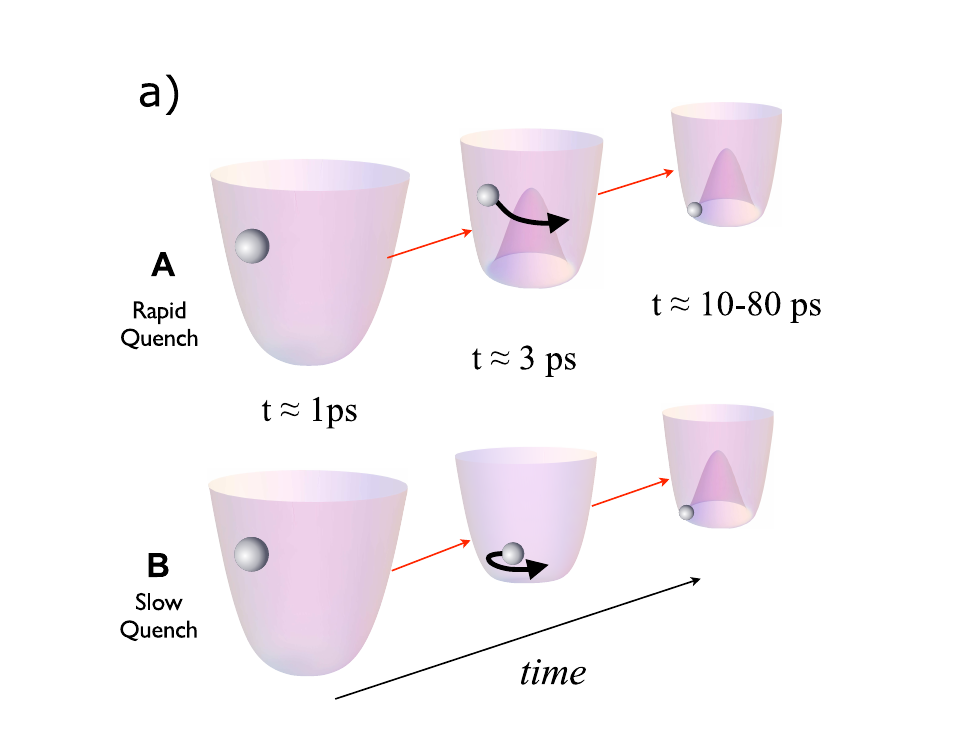}
\par\end{center}

\includegraphics[width=1\columnwidth]{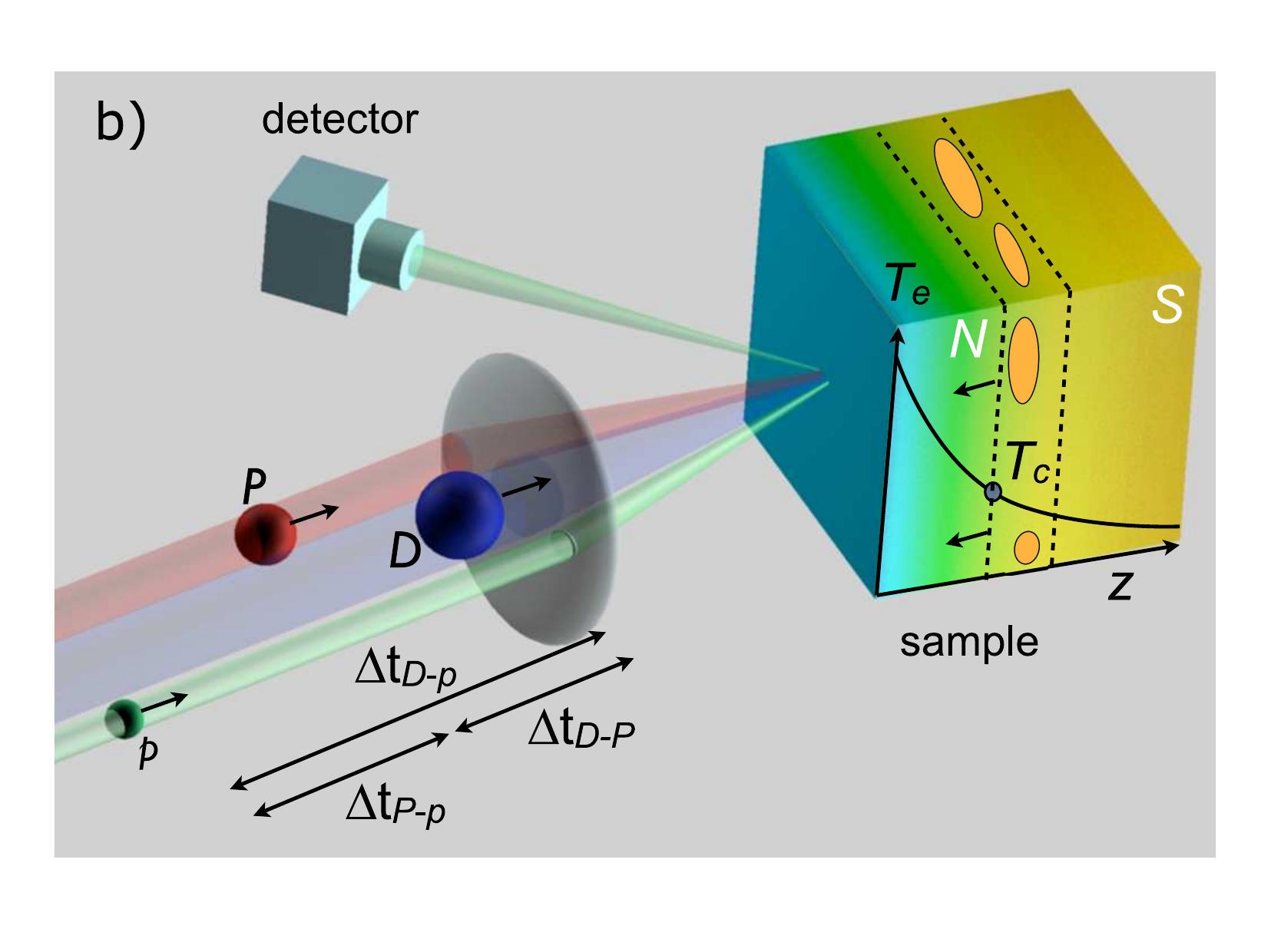}%
\end{minipage}
\par\end{centering}

\centering{}\protect\caption{a) The system trajectory (depicted by the silver ball) in a temporally
evolving potential. In the rapid quench scenario (\textbf{A}), the
potential changes faster than the system can follow. The opposite
is true in the slow quench scenario (\textbf{B}). b) A schematic diagram
of the pulse sequence. The time delays $\Delta t_{\mathrm{D-P}},\Delta t_{\mathrm{D-pr}}$
and $\Delta t_{\mathrm{P-pr}}$ refer to delays between the \emph{D,}
\emph{P} and \emph{pr} pulses depicted in blue, red and green respectively.
The $S/N$ phase boundary moves with velocity $v_{\psi}$ towards
the surface. Vortices are created in the wake of the temperature front
whose position is given by $T(r,t)=T_{\mathrm{c}}$.}
\centering{}\label{fig1} 
\end{figure}

\subsection*{Experimental results}

To measure the trajectory of the system through the $N\rightarrow S$
transition, we use a 3 pulse technique shown schematically in Fig.
1 b).  (See also Supplementary Information). The first, destruction
(\emph{D}) laser pulse strongly perturbs electronic subsystem initiating
$S\rightarrow N$ transition on the timescale of $\sim0.8$ ps.\cite{Kusar2008}
The recovery of the $S$ state in the ensuing $N\rightarrow S$ transition
is measured by means of the pump-probe (\emph{P-pr}) transient reflectivity
$\Delta R(\Delta t_{\mathrm{P-pr}})/R$ measurements. The pump-probe
response is recorded at a set of delays $\Delta t_{\mathrm{D-P}}$
between \emph{D} and \emph{P} pulses\cite{Yusupov2010}. For each
value of the $\Delta t_{\mathrm{D-P}}$ delay the amplitude of the
response $A=(\Delta R/R)^{max}$ is extracted, and, when plotted as
a function of  $\Delta t_{\mathrm{D-P}}$, is a measure of the trajectory
of the system. It is then compared to the modeled behavior of the
order parameter $\psi(t)$ using an appropriate response function
(See Supplement for a rigorous discussion).

We present measurements on a La$_{1.9}$Sr$_{0.1}$CuO$_{4}$ (LSCO)
single crystal as a prototype single-layer cuprate well studied by
means of pump-probe technique.\cite{Kusar2005,Kusar2008} The intermediate
value of the $T_{\mathrm{c}}=28$ K is relatively high so that systematic
with fluence experiments can be conducted, and simultaneously it is
low enough so that the theoretical estimate of $\tau_{\mathrm{GL}}\approx100$
fs is longer than our resolution (30 fs). The laser fluence required
to destroy superconducting state on the surface (photodestruction
threshold) has been previously determined to be $\mathcal{F}_{\mathrm{T}}=4.2\mathrm{\pm1.7\mbox{ }\mu J/cm^{2}}$.\cite{Kusar2008}
In the presented experiment we vary the \emph{D-}pulse fluence from
$\mathcal{F}_{\mathrm{T}}$ up to 34 $\mu\mbox{J/c\ensuremath{m^{2}}}$. 

\begin{figure}[t]
\includegraphics[width=1\columnwidth]{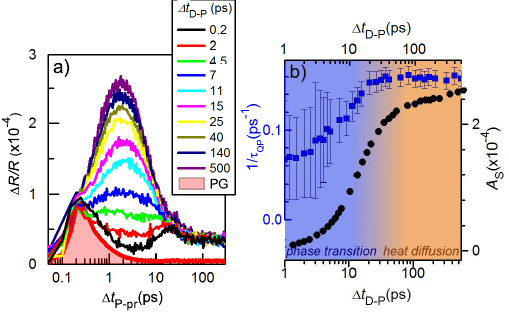}

\protect\caption{a) The transient reflectivity $\Delta R/R$ for La$_{1.9}$Sr$_{0.1}$CuO$_{4}$
at 4 K as a function of $\Delta t_{\mathrm{P-pr}}$, for different
$\Delta t_{\mathrm{D-P}}$. The D pulse fluence is $\mathcal{F}_{\mathrm{D}}=12\:\mu$J/cm$^{2}$,
which is approximately three times above the destruction threshold
($\mathcal{F}_{\mathrm{T}}\simeq4\mu$J/cm$^{2}$)\cite{Kusar2008}.
The red line indicates the pseudogap signal measured above $T_{\mathrm{c}}$.
b) Black squares - the amplitude of the superconducting component
$A_{\mathrm{S}}$ extracted from a) after subtraction of the PG.\textcolor{black}{{}
The blue squares are $1/\tau{}_{\mathrm{QP}}$ as a function of $\Delta t_{\mathrm{D-P}}$.
The recovery of the system is schematically divided into phase transition
region (blue background) where order parameter is not thermal, and
into the thermal diffusion (orange background) where the transition
is effectively over and order parameter is defined solely by the temperature.\label{fig:RawData}}}
\end{figure}

A representative dataset obtained in a three-pulse experiment is shown
in Fig. \ref{fig:RawData}a). It depicts the transient reflectivity
$\Delta R(\Delta t_{\mathrm{P-pr}})/R$ traces for different $\Delta t_{\mathrm{D-P}}$
delays during the system recovery measured at 4K with $D$ pulse fluence
$\mathcal{F}_{\mathrm{D}}=12\mathrm{\mbox{ }\mu J/cm^{2}}$. Two distinct
and easily identifiable contributions are observed: a pseudogap (PG)
response $(\Delta R/R)_{\mathrm{PG}}$ which peaks around 0.2 ps,
and the quasiparticle (QP) recombination across the superconducting
gap, i.e. the superconducting response $(\Delta R/R)_{\mathrm{S}}$
which peaks near 2 ps, extending to tens of ps.\cite{Kusar2010,Kusar2005}
We clearly observe that  $(\Delta R/R)_{\mathrm{S}}$ gradually increases
with increasing delay $\Delta t_{\mathrm{D-P}}$ indicating the recovery
of the S state, while $(\Delta R/R)_{\mathrm{PG}}$ remains intact
by the destruction pulse and does not show any change with $\Delta t_{\mathrm{D-P}}$.

$(\Delta R/R)_{\mathrm{PG}}$ is known to be independent of \emph{T}
 at temperatures below 100 K in this material.\cite{Kusar2005} We
are interested in the superconducting order, so for further analysis
we subtract it from the data %
\footnote{ For maximum accuracy we subtract signal at $\Delta t_{\mathrm{D-P}}=0.2$
ps after the \emph{D} pulse at 34 K (6 K above $T_{\mathrm{c}}$ ).
This allows to completely separate pseudogap from superconducting
signal and give a good estimate for remaining signal at timescales
up to 1 ps.%
} and plot \textcolor{black}{the amplitude} of the remaining \textcolor{black}{superconducting
response $A_{\mathrm{S}}$} as a function of $\Delta t_{\mathrm{D-P}}$
in Fig. 2 b)\textcolor{black}{{} (shown in black circles). }

\textcolor{black}{From the exponential fits of the initial decay of
$(\Delta R/R)_{\mathrm{S}}$}$(\Delta t_{\mathrm{P-pr}})$\textcolor{black}{{}
we obtain the QP relaxation time $\tau_{\mathrm{QP}}$ as a function
of $\Delta t_{\mathrm{D-P}}$ plotted in Fig \ref{fig:RawData} b).}%
\footnote{For the analysis of the relaxation time the dataset at \textcolor{black}{$\Delta t_{\mathrm{D-P}}=0$}
is subtracted. This allows to remove the long-lived contribution after
$\tau_{\mathrm{th}}$, which originates in pump induced heating and
is independent on $\Delta t_{\mathrm{D-P}}$%
}\textcolor{black}{{} We observe that 1/$\tau_{\mathrm{QP}}$ shows a
similar to $A_{\mathrm{S}}$ time-evolution. If we assume that $1/\tau_{\mathrm{QP}}\propto\Delta_{\mathrm{S}}$,
where $\Delta_{\mathrm{S}}$ is the superconducting gap\cite{Kabanov1999,Schuller2006},
the observed dependence of $\tau_{\mathrm{QP}}$ on }$\Delta t_{\mathrm{D-P}}$\textcolor{black}{{}
is consistent with opening of the S gap for $\Delta t_{\mathrm{D-P}}\sim0$.
The two variables $A_{\mathrm{S}}$ and $\tau_{\mathrm{QP}}$ identify
the recovery of superconducting order on a 10 ps timescale. The measured
dependence of the trajectory $A_{\mathrm{S}}(\Delta t_{\mathrm{D-P}})$
for different fluences $\mathcal{F}$ is shown in Fig. \ref{dest-rec},
where it is compared to the simulated trajectories from different
models described below.}

\subsection*{Modeling with time-dependent Ginzburg-Landau theory}

In following we try to formulate the minimal  TDGL theory which captures
the observed behavior using laser pulse fluence as the only variable
parameter. We consider only the real part of TDGL equations as the
optical response is insensitive to the phase of the order parameter.\cite{Madan2014}
(In the supplement we solve the full set of TDGL equations to qualitatively
account for dynamics of the phase and vortex dynamics.) The basic
TDGL equation describing the order parameter $\psi(t,z)$ dynamics
is \cite{Gorkov1975}: 
\begin{equation}
\frac{\partial\psi}{\partial t}=\alpha_{r}(t,z)\psi-\psi|\psi|^{2}+\nabla^{2}\psi,\label{eq:tdgl}
\end{equation}
where we have omitted explicit dependence of $\psi$ on $t$ and $z$,
and the temporal and spatial coordinates are measured in units of
$\tau_{\mathrm{GL}}$(fitting variable) and coherence length ($\xi=0.2$
nm\cite{Suzuki1991}) at $T=0$ K, respectively.

The system is driven by the electronic temperature $T_{\mathrm{e}}$,
which enters TDGL via $\alpha_{\mathrm{r}}(t,z)=(1-T_{\mathrm{e}}(t,z)/T_{\mathrm{c}})$.
The temperature is time dependent, and also depends on the depth in
the sample. To calculate the actual shape of $T_{\mathrm{e}}(t,z)$,
we assume that electrons are preferentially coupled to a particular
boson (phonon and/or spin excitation), which in turn releases its
energy to the lattice. This three-temperature model (3TM) has been
used in the past to describe the normal state ultrafast response in
unconventional superconductors. \cite{Perfetti2007,Mansart2010d}
  In principle, the 3TM describes the destruction of the condensate,
defines the recovery timescales and should also describe the slow
diffusion processes clearly present in the data (Fig. \ref{fig:RawData}
b). To account for the latter we introduce thermal diffusivity $\kappa$
as a fitting parameter, which does not affect short timescales. The
final set of  the equations from which we obtain $T_{\mathrm{e}}(t,z)$
is then: 
\begin{align}
\gamma_{\mathrm{e}}T_{\mathrm{e}}\dot{T_{\mathrm{e}}}=-\gamma_{\mathrm{ep}}(T_{\mathrm{e}}-T_{\mathrm{p}})+P(t)\nonumber \\
C_{\mathrm{p}}\dot{T_{\mathrm{p}}}=-\gamma_{\mathrm{ep}}(T_{\mathrm{p}}-T_{\mathrm{e}})-\gamma_{\mathrm{pl}}(T_{\mathrm{p}}-T_{\mathrm{l}})\\
C_{\mathrm{l}}\dot{T_{\mathrm{l}}}=-\gamma_{\mathrm{pl}}(T_{\mathrm{l}}-T_{\mathrm{p}})+\kappa\frac{\partial^{2}T_{\mathrm{l}}}{\partial z^{2}}.\nonumber 
\end{align}

where $\gamma_{\mathrm{e}}=2.5\mbox{mJ/mol/}\mbox{K}^{2}$ \cite{Wen2004b}
is electronic specific heat coefficient, $\gamma_{\mathrm{ij}}$ represents
the coupling between the $i$th and $j$th bath, $T_{\mathrm{i}}$
is the temperature of the corresponding system (indices e, p and l
are for electronic, hot boson and lattice, respectively), and $P(t)=\frac{\mathcal{F}}{2\pi\omega}\exp{(-t/2w^{2})}\exp{(-z/\lambda)}$
is the optical excitation, $w=60\, fs$ is the pulse width at half
maximum. We also assume that total phonon heat capacity $C=C_{\mathrm{p}}+C_{\mathrm{l}}$,
where $C_{\mathrm{p}}=\alpha C_{0}$ and $C_{\mathrm{l}}=(1-\alpha)C_{0}$
is heat capacity of hot bosons and the lattice bath, respectively,
$\alpha=0.2$ is a fraction of the phonons modes which are strongly
coupled to electrons. \cite{Perfetti2007}  The coefficients of the
3TM model $\gamma_{\mathrm{ep}}$ and $\gamma_{\mathrm{pl}}$ can
be estimated using electronic and lattice thermal constants $\gamma_{\mathrm{e}}$
and $C$, measured electron-phonon relaxation rate $\gamma_{\mathrm{l}}=340\mbox{K/ps}$
and the phonon-phonon relaxation time $\tau_{\mathrm{ph}}=0.6\mbox{ ps}$
\cite{Gadermaier2010}: $\gamma_{\mathrm{ep}}=\gamma_{\mathrm{e}}\gamma_{\mathrm{l}}$
and $\gamma_{\mathrm{pl}}=C_{\mathrm{p}}/\tau_{\mathrm{ph}}$. The
temperature dependence of the phonon heat capacity $C$ is obtained
from published thermal data.\cite{Junod1987b}

\begin{figure}
\includegraphics[width=0.9\columnwidth]{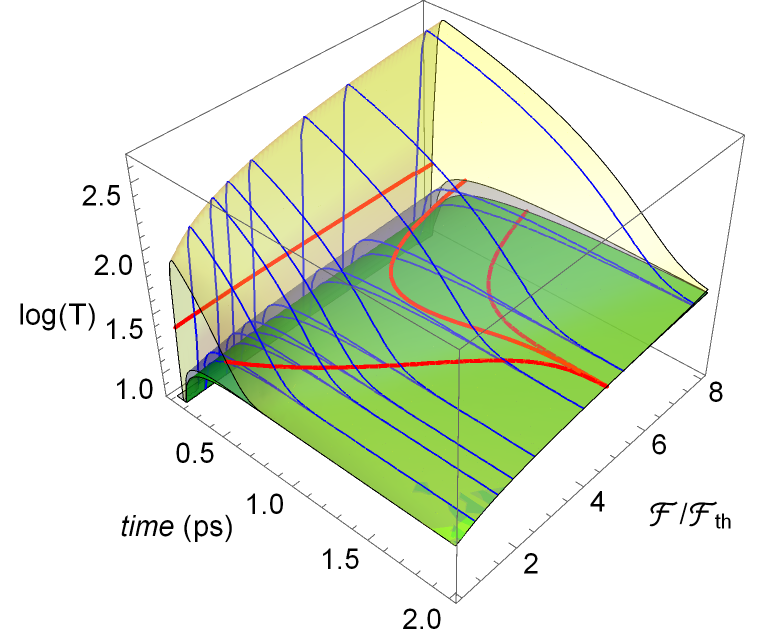}

\protect\caption{Three surfaces showing the calculated time evolution of the logarithm
of $T_{\mathrm{e}}$ (yellow), $T_{\mathrm{p}}$ (blue) and $T_{\mathrm{L}}$
(green) as a function of incident fluence. The initial temperature
of the sample is 4 K. $T_{\mathrm{p}}$ and $T_{\mathrm{L}}$ are
very close, but $T_{\mathrm{e}}$ reaches in excess of 400 K. The
blue lines correspond to fluences used in the experiment. The red
lines indicate where $T_{\mathrm{e}}$ , $T_{\mathrm{p}}$ and $T_{\mathrm{L}}$
cross $T_{\mathrm{c}}=28$ K. $T_{\mathrm{e}}$ is used in the modeling
of the order parameter. Fast quench corresponds to $\mathcal{F}<5\,\mathcal{F}_{\mathrm{th}}$$\simeq23\,\mu$J/cm$^{2}$.}
\label{3TM-fig}
\end{figure}

Solving Eq. (2) we obtain the time and depth dependence of $T_{\mathrm{e}}$,
$T_{\mathrm{p}}$ and $T_{\mathrm{l}}$. In Fig. \ref{3TM-fig} we
plot the values of corresponding temperatures on the sample surface
for different fluences used in the experiments. Initially, the pulse
rapidly heats the electronic system, but energy is quickly transferred
to the strongly coupled bosons and the lattice on a timescale $\sim1$
ps, whereafter the three temperatures rapidly merge. Note that this
timescale is of the same order as the destruction of the $S$ state.\cite{Stojchevska2011}
We see in Fig. \ref{3TM-fig} that for low excitation fluences $\mathcal{F}/\mathcal{F}_{\mathrm{th}}$,
the quench rate $\gamma_{\mathrm{Q}}=(dT_{e}/dt)_{T_{\mathrm{c}}}$
through $T_{\mathrm{c}}$ (red line) is fast, on the order of $4\times10^{14}$
K/s and $T_{\mathrm{e}}\simeq T_{\mathrm{p}}\simeq T_{\mathrm{l}}$
already after $\sim$ 1 ps. With higher fluences, $>5$ $\mathcal{F}_{\mathrm{th}}$
when $T_{\mathrm{p}}$ and $T_{\mathrm{l}}$ both exceed the superconducting
$T_{\mathrm{c}}$, the cooling rate is mainly determined by thermal
diffusion on timescales well beyond $\sim1$ ps. We emphasize that
this cross-over from rapid to slow quench is quite general and does
not rely on the specific details of the 3TM. Having calculated $T_{\mathrm{e}}(t,z)$,
we can now calculate $\psi(t,z)$, and $A_{\mathrm{S}}(t)$.

\begin{figure*}[tbh]
\includegraphics{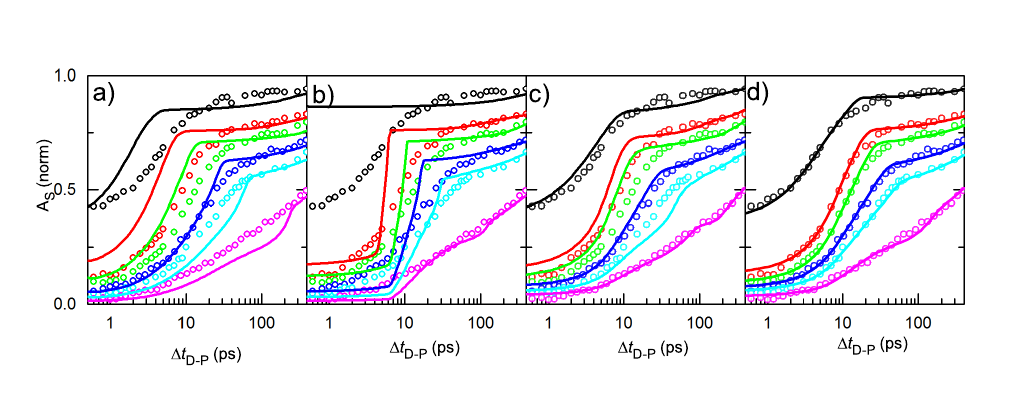}\protect\protect\caption{\label{Fig_Coeff} Comparison of experimentally measured data (circles,
values of fluence are 4 (black), 9 (red), 12 (green), 18 (blue), 24
(cyan), and 34 (magenta) $\mu\mathrm{J/cm^{2}}$) and the calculated
$A_{S}$ obtained within different formulation of problem (solid lines):
TDGL solution with initial conditions described by a) $\psi(0,z)=\sqrt{1-T_{\mathrm{bath}}/T_{\mathrm{c}}}$,
b) Eq. \ref{eq:IniCond}, c) Eq. \ref{eq:EqVolovik} (represented
in Fig. \ref{fig:IC}), d) Eq. $\psi_{\mathrm{fluc}}(0,z)=\kappa z$.}

\label{dest-rec} 
\end{figure*}

Within the \textbf{first approach} to the problem presented above,
Eq. (\ref{eq:tdgl}) describes \emph{both} the destruction and recovery
of the superconducting condensate. First we consider in detail the
destruction stage, i.e. $S\rightarrow N$ transition. Initial conditions
describe superconducting state in equilibrium $\psi(0,z)=\sqrt{1-T_{\mathrm{bath}}/T_{\mathrm{c}}}$.
The transition is then driven by a temperature burst described by
the $\alpha$-term of Eq. (\ref{eq:tdgl}). Within this approach the
relation between $\tau_{\mathrm{GL}}$ (the only free parameter) and
duration of the temperature burst is crucial. For the condensate to
be able to react to the intense short temperature perturbation $\tau_{\mathrm{GL}}$
should be smaller than duration of perturbation $\sim1$ ps. Such
short $\tau_{\mathrm{GL}}$ results also in rapid recovery, significantly
faster than observed experimentally, as can be seen in Fig. \ref{dest-rec}
a) ($\tau_{\mathrm{GL}}=450$ fs ). At fluences $\mathcal{F}\geq18\,\mu\mathrm{J/cm^{2}}$
we enter the slow quench regime, i.e. perturbation duration is longer
and OP suppression is more effective. This results in a recovery that
occurs on the time-scale closer to experimentally observed. 

On the other hand, if $\tau_{\mathrm{GL}}$ is much longer than the
perturbation, the condensate cannot follow the temperature dynamics
and the condensate remains undestroyed. Thus the presented TDGL equations
do not provide a good description of the destruction of the condensate.
This can be easily understood as the destruction is the fastest of
considered processes during which the distribution function is clearly
not thermal  which leads to effects which lie beyond a TDGL and 3TM
description.\cite{Baranov2014}

To avoid the problem of the condensate photodestruction we focus on
the recovery process, and define the state of the system after the
$S\rightarrow N$ transition by the initial conditions. The solution
of the TDGL equation then describes the ensuing recovery dynamics,
i.e. the $N\rightarrow S$ transition. Within our \textbf{second approach}
we determine initial depth-distribution of the condensate density from
the fluence dependence of the amplitude in two pulse pump-probe response
 experiments\cite{Kusar2008}: 

\begin{equation}
\psi(0,z)=\begin{cases}
0\! & \!,\mathcal{F}(z)>\mathcal{F}_{\mathrm{T}};\\
(1-\frac{\mathcal{F}}{\mathcal{F}_{\mathrm{T}}}e^{-z/\lambda})\sqrt{1-\frac{T_{\mathrm{l}}(0,z)}{T_{\mathrm{c}}}}\! & \!,\mathcal{F}(z)<\mathcal{F}_{\mathrm{T}}.
\end{cases}\label{eq:IniCond}
\end{equation}


This expression can be considered as the limiting case of the fast
quench without fluctuations. The solution of the TDGL equations then
corresponds to an $S\to N$ boundary which moves towards the surface
in the form of a $S/N$ soliton wall without change of shape, and
the recovery of the system is completely determined by the soliton
propagation. This approach has only one free fit parameter $\tau_{\mathrm{GL}}$
defining the velocity of the soliton $v_{\mathrm{s}}$\prettyref{fn:SOliton}.
By setting $\tau_{\mathrm{GL}}=50\,\mathrm{fs}$ we obtain recovery
on observable timescales. However, the obtained trajectories $A_{\mathrm{S}}(\Delta t_{\mathrm{D-P}})$
are much sharper than experimentally observed (Fig. \ref{dest-rec}
b). For the weakest excitation $\mathcal{F}(z=0)=\mathcal{F}_{\mathrm{T}}=4.2\:\mu\mathrm{J/cm^{2}}$
the condensate is destroyed only at the surface and recovery occurs
in $50\,\mathrm{fs}$. However, for the strongest excitation $\mathcal{F}=34\:\mu\mathrm{J/cm^{2}}$
the slow quench condition is satisfied and the soliton follows the
temperature front. In this case the simulated curve fits reasonably
well to the data (Fig. \ref{dest-rec} b).

The gradual growth of the $S$ signal on the 10 ps timescale for the
intermediate fluences might be reproduced if one assumes that the
main source of the emerging order is superconducting fluctuations
along the Kibble-Zurek scenario. It has been widely discussed that
cuprates have extremely strong fluctuations compared to conventional
superconductors and their onset in certain families exceeds the critical
temperature by several tens of K above $T_{\mathrm{c}}$.\cite{Xu2000d,Madan2014,Junod2000,Corson1999,Perfetti2015}
Thus we expect to have a significant contribution from fluctuations
and expect them to be an effective seed for the order parameter growth. 

In a simple pre-formed pair scenario of superconductivity, the pseudogap
energy corresponds to the pair formation energy. One thus can assume
that superconductivity arises from pseudogap-forming carriers. Their
density after the photoexcitation can be described by equation \ref{eq:IniCond},
where $\mathcal{F_{\mathit{\mathrm{T}}}}$ would now  correspond to
the pseudogap photodestruction fluence $\mathcal{F_{\mathit{\mathrm{PG}}}}$.
In LSCO this fluence is extremely high $\mathcal{F_{\mathit{\mathrm{PG}}}}\sim750\pm200$$\mu$J/cm$^{2}$.\cite{Kusar2010}
Under excitation conditions considered in this work $\mathcal{F}\leq34\:\mu\mathrm{J/cm^{2}}$
pseudogap would hardly be affected at all, which is confirmed by the
robustness of the PG response in three-pulse experiments. This implies
that the initial conditions for equation \prettyref{eq:IniCond} would
be more or less constant with depth and fluence. Such fluence-independent
initial conditions are inconsistent with the fluence-dependence of
the data, so the PG does not appear to seed the $S$ order parameter,
and does not improve the model fit.

\begin{figure}
\includegraphics[width=1\columnwidth]{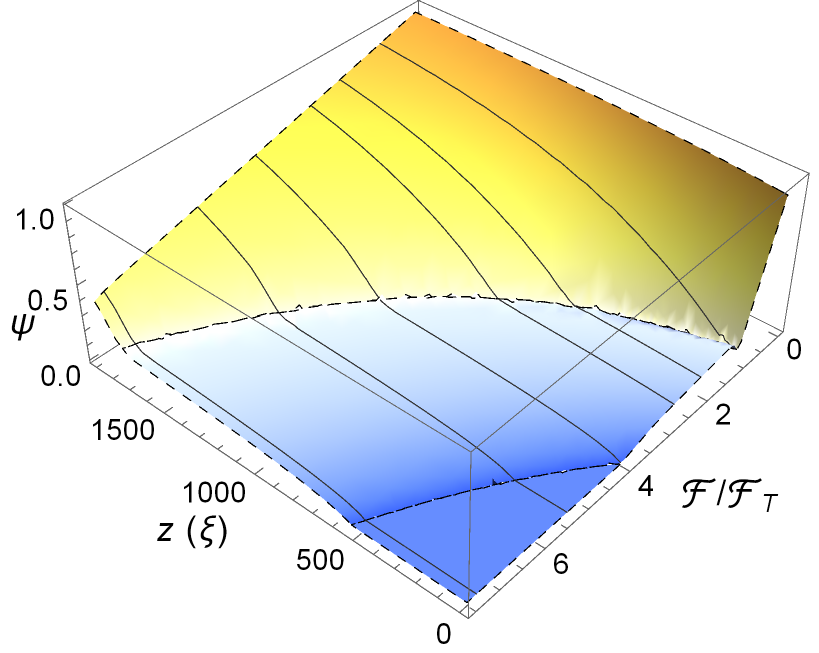}

\protect\caption{Initial condition that take into account fluctuations of the order
parameter according to the Eq. \prettyref{eq:EqVolovik} (blue surface)
and partially suppressed (yellow surface) order parameter in the region
$\mathcal{F}(z)<\mathcal{F}_{\mathrm{T}}.$ in agreement with Eq.
\prettyref{eq:IniCond}.\label{fig:IC}}
\end{figure}

\textbf{The third approach: thermal fluctuations.} A more relevant
scenario invokes  fluctuations of the superconducting order above
$T_{\mathrm{c}}$. After the $S$ order has been destroyed and the
temperature is still significantly above $T_{\mathrm{c}}$ ($\Delta t_{\mathrm{D-P}}$$\sim0.2$
ps) weak short-lived superconducting fluctuations exist in the system.
As the electronic subsystem rapidly cools  the density of the $S$
fluctuations increases, and their lifetime and correlation length
diverge as $T\rightarrow T_{\mathrm{c}}$. During the initial cooling
stage $(T\gg T_{\mathrm{\mathrm{c}}})$ fluctuations are fast and
 would adapt to the variation in temperature. However, after a certain
moment in time given by $\tau_{\mathrm{Z}}=\sqrt{\tau_{\mathrm{GL}}\tau_{\mathrm{q}}}$
their lifetime would become larger than the quench time, meaning that
the system would cross the transition in a \textquotedbl{}frozen\textquotedbl{}
inhomogeneous configuration. As the system escapes the critical region
$T<T_{\mathrm{c}}(1-\frac{\tau_{\mathrm{\mathrm{Z}}}}{\tau_{\mathrm{q}}})$
the fluctuation lifetime decreases and the system starts to adapt
to the new conditions, i.e. the order parameter grows from fluctuations
according to TDGL theory. The appropriate  expression for such fluctuations,
which we can implement as initial conditions have been given by Volovik
\cite{Volovik2000}%
\footnote{In the original paper the misprint occurred - the power of the fraction
was 1/8 instead of 3/8%
}
\begin{equation}
\psi_{\mathrm{fluc}}(0,z)\sim\left(\frac{\tau\mbox{\ensuremath{_{\mathrm{GL}}}}}{\tau_{\mathrm{q}}(z)}\right)^{3/8}\frac{T_{\mathrm{c}}}{E_{\mathrm{F}}}\psi_{\mathrm{eq}}(T_{\mathrm{l}}).\label{eq:EqVolovik}
\end{equation}
 Factor $\frac{T_{\mathrm{c}}}{E_{\mathrm{F}}}$ gives the correct
order of magnitude $\sim0.01$ of the seed order parameter, whereas
$\tau_{\mathrm{q}}$ and $\psi_{\mathrm{eq}}(T_{\mathrm{l}})$ define
the depth dependence. The spatial dependence of the initial order
parameter is shown in Fig. \ref{fig:IC}. We note that the actual
initial temperature after the photon absorption is not important because
the properties of the seed OP are defined at $T_{\mathrm{Z}}=T_{\mathrm{c}}(1+\frac{\tau_{\mathbb{Z}}}{\tau_{\mathrm{q}}})$.
 For the initial condition of the equation \prettyref{eq:tdgl} we
use the equation \prettyref{eq:EqVolovik} with the proportionality
factor $C$: $\psi_{\mathrm{ini}}\left(\mathcal{F}(z)>\mathcal{F}_{\mathrm{T}}\right)=C\cdot\psi_{\mathrm{fluc}}(0,z)$
which, together with $\tau_{\mathrm{GL}},$ was used as an adjustable
parameter shared between all curves. The resulting trajectories are
shown in Fig. \ref{dest-rec} c) (parameter values are $\tau_{\mathrm{GL}}=1.25$
ps and $C=4$). The agreement between experimental data and simulations
is now much better. 

Finally, we improve the fit by  solving the TDGL equation with parametrized
phenomenological initial conditions which resemble main feature of
the Volovik's result i.e. growth of the fluctuations with depth. The
goal here is to provide initial conditions where depth dependence
is adjustable rather than defined by the quench-rate deduced from
the 3TM which may not be sufficiently accurate. Instead of equation
\prettyref{eq:EqVolovik} we use the minimal model which produces
a reliable fit: $\psi_{\mathrm{fluc}}(0,z)=\kappa z/\lambda_{\mathrm{p}}$,
where $\kappa$ is the fitting parameter independent for each fluence
value. Result of this simulations is presented in Fig. \ref{dest-rec}
d) with $\kappa$ values equal to 0.25, 0.23, 0.162, 0.13, 0.1 and
0.06 in the order of the increasing fluence, and $\tau_{\mathrm{GL}}=1.1$
ps. The good agreement of the simulation with the data justifies the
initial conditions within the fluctuation scenario and underlines
importance of the fluctuations especially for low and intermediate
fluences. 

We conclude that the evolution of the superconducting order through
the non-equilibrium phase transition can be described quite well with
time-dependent Ginzburg Landau theory. We show in the supplement that
the remaining discrepancy between the fit and the data near $\sim10$
ps for the two fastest quench rates can be accounted for by the suppression
of the order parameter due to vortex formation. \textbf{ }From the
discussion above one can see that in the case of a fast quench the
ergodicity of the system is broken   as soon as the fluctuation timescale
becomes longer than the quench time.  The system then cannot follow
the time-evolution of the potential  and so evolves inhomogeneously
through the transition with the creation of vortices, quite faithfully
reproduced by the model when the dynamics of the superconducting phase
is explicitly included in the calculations (Fig. 4 of the supplement). 

The fact that the \emph{destruction} of the condensate cannot be properly
described by TDGL equations is not surprising, particularly if we
rely only on the 3TM, which does not take into account the details
of the destruction process. On the other hand, in spite of all its
inherent shortcomings, the ensuing recovery of \emph{S} order predicted
by the model  agrees quite well with the experiments, and emphasizes
a crucial role of fluctuations in the time-evolution of the order
parameter through the $N\rightarrow S$ transition. Remarkably, the
experiments show that the $S$ order appears to grow out of the pseudogap
state, yet from the strong fluence dependence of the\emph{ S} recovery
and complete fluence-independence of the \emph{PG} state we conclude
that the pseudogap has no effect in seeding the emergence of the superconducting
order. 
\begin{acknowledgments}
We wish to acknowledge the useful discussion with T.W. Kibble regarding
the importance of a variable quench rate in the experiment. The funding
was provided by European Research Council advanced grant TRAJECTORY.
\end{acknowledgments}

\bibliographystyle{apsrev4-1}

\end{document}